# CBCT-IQ: A Publicly Available Annotated Cone-Beam CT Dataset for Image Quality Assessment and Benchmarking

Sepideh Hatamikia[1,2,3*], Anna Breger[3,4], Clemens Karner[3,2], Birgit Pohn[2,5], Poorya MohammadiNasab[2,3], Martin Buschmann[6], Stephanie Nougaret[7], Laura Haddad[7], Ali Abbasian Ardakani[2], Afshin Mohammadi[8], Paul Apfaltrer[9], Wolfgang Birkfellner[3], Alfred Pohl[2,10], Ander Biguri[4], Gernot Kronreif[1], Carola-Bibiane Schönlieb[4], Tess Reynolds[11]

[1]Austrian Center for Medical Innovation and Technology (ACMIT), Wiener Neustadt, Austria

[2] Department of Medicine, Faculty of Medicine and Dentistry, Danube Private University, Krems, Austria

[3] Center for Medical Physics and Biomedical Engineering, Medical University of Vienna, Vienna, Austria

[4] Department of Applied Mathematics and Theoretical Physics (DAMTP), University of Cambridge, Cambridge, United Kingdom

[5] Faculty Computer Science and Applied Mathematics, University of Applied Sciences Technikum Wien, Vienna, Austria

[6] Department of Radiation Oncology, University Hospital Vienna and Medical University of Vienna, Vienna, Austria

[7] Department of Radiology, Montpellier Cancer Institute, PINKCC lab U1194, University of Montpellier, Montpellier, France

[8] Department of Radiology, Faculty of Medicine, Urmia University of Medical Science, Urmia, Iran

[9] Clinical Institute for Diagnostic and Interventional Radiology and Nuclear Medicine, University Hospital Wiener Neustadt, Wiener Neustadt, Austria

[10] Clinical Department of Ophthalmology and Optometry, University Hospital Wiener Neustadt, Wiener Neustadt, Austria

[11] Image X Institute, Faculty of Medicine and Health, University of Sydney, Australia

*** Corresponding author: Sepideh Hatamikia, E-mail:** Sepideh.hatamikia@DP-Uni.ac.at

**Abstract:**

Medical image quality plays a critical role in diagnostic accuracy, especially in X-ray–based imaging modalities such as cone-beam computed tomography (CBCT), where image quality must be balanced against radiation dose. While expert visual evaluation remains the clinical standard for image quality evaluation, it is time-consuming, subjective and affected by inter-observer variability, emphasizing the need for reliable quantitative image quality assessment (IQA) methods. However, the development and validation of such IQA methods have been limited by the lack of publicly available CBCT datasets with expert image quality annotations. In this study, we provide the first open-access CBCT IQA dataset containing 1,764 annotated image slices acquired using systematic variations in image acquisition and reconstruction parameters. Three clinical experts graded the overall image quality and a predefined regions of interest (ROI) using a four-level scoring scheme. In addition, we benchmark 26 full reference- and no reference-based IQA measures against expert annotations and introduce an exploratory IQA measure–based ranking capable of distinguishing subtle image quality differences. This dataset introduced a standardized benchmark for future CBCT IQA research and provides a valuable resource for the development and validation of new IQA methods, enabling reproducible research and advancing CBCT IQA.

## Background and Summary

The quality of medical images is a critical factor influencing diagnostic accuracy [1]. In X-ray–based modalities such as computed tomography (CT) and cone-beam CT (CBCT), where there is always a trade-off between radiation dose and image quality, image quality assessment (IQA) can be a key factor in radiation dose optimization [2, 3]. Image quality assessment can be performed manually by expert radiologists and clinicians, which provides clinically meaningful evaluations. However, this approach has several drawbacks, including inter-observer variability, especially in complex medical scenarios, as well as being time-consuming and costly. Therefore, there is a need for standardized quantitative IQA measures that provide an image quality score with a high correlation to human readers [4, 5].

Unlike conventional CT, which utilizes a linear detector array and requires many rotations around the patient to acquire volumetric data, CBCT employs a 2D detector and a cone-shaped X-ray beam, allowing it to capture the entire volume in a single rotation. CBCT generally exposes patients to a lower radiation dose than conventional CT while providing volumetric imaging. In recent years, CBCT has gained considerable attention in various applications, including dentistry [6, 7], orthopedics [8], angiography [9], and image-guided surgery [10, 11]. Given its growing clinical relevance, rigorous evaluation of emerging CBCT technologies and reconstruction algorithms using appropriate quantitative IQA measures is essential [12]. Therefore, the development and identification of robust and reliable IQA measures capable of accurately quantifying CBCT image quality are of critical importance in this context.

Furthermore, development of accurate quantitative IQA measures is indispensable in the development stage of machine learning (ML) algorithms that operate on medical images [13, 14]. As opposed to non-ML methods in imaging, where explainability of the method can be of use for evaluating their value and performance, the black-box nature of most ML methods inherently requires that their evaluation must be done purely empirically, by testing their quality via IQA measures. This is doubly true in reinforcement learning, where the loss (or policy) can be any measure, including IQA. Due to this empirical need in the ML community IQA has become, to a degree, the gold standard for evaluating the methods contribution, and thus critical in the success of these methods into clinic [15-18].

However, the complexity of quantifying an image quality to be approximated by an IQA measure and the variety of possible algorithm designs for IQA is a major difficulty. Thus, understanding well what different existing IQA measures, when they fail, what type or artifacts they promote or ignore is fundamental not only for good scientific approaches, but also a critical point in the age of ML methods, where IQA is often the only parameter to decide on which method is most valid in an application.

IQA measures can be generally divided into no reference–based and full reference–based methods. No reference–based measures are more applicable in real clinical cases, where the gold-standard image is usually unavailable, while reference-based methods show a higher correlation with human expert ratings, as they compare the image to a known high-quality ground truth [13, 14]. In addition, full-reference measures are more useful in the development process of new methods, where evaluating performance with respect to a known high-quality image is applicable.

**CBCT imaging and IQA**

Despite the existence of many classical and modern IQA measures [19, 20], many studies argue that there is a gap of efficient IQA for medical images also including CBCT images [21, 22] where this discrepancy will lead to incorrect evaluations of devices and algorithms. Therefore, the evaluation and development of robust and efficient IQA measures for CBCT data remain vital and are still insufficiently explored. The development of quantitative IQA measures requires initial validation by clinical experts to ensure a strong correlation with their subjective perception of medical image quality.

The main challenge is that such annotations are not easy to obtain. Researchers working on the development of quantitative IQA measures regularly face the problem of limited access to datasets in which image quality has been graded by expert clinicians. Recently, to tackle this challenge, there have been first data sets published that provide medical image data and corresponding expert quality ratings, for photoacoustic and full-reference data [23] and for CT in the no-reference setting [24]. However, following these directions, there is a clear need for a comprehensive and representative dataset of CBCT images with corresponding annotated image quality levels, which would be essential to addressing this challenge.

In this work, the first open-access CBCT dataset specifically designed to provide images along with clinical annotations of image quality levels is presented. The dataset is composed of systematic variations in acquisition parameters, such as kV, pulse width as well as reconstruction types available on the Siemens ARTIS pheno imaging system, thereby producing a spectrum of image quality levels ranging from poor to high image quality. Three clinical experts performed annotations on different 2D image slices of each 3D CBCT volume providing scores classified in four image quality levels. In addition to the entire image quality label, the assessment of defined region of interest (ROI) is also performed for each 2D image. Two anatomical phantoms from thorax and pelvis were used for all different acquisition setups.

Annotators were shown ground truth images next to corresponding degraded images and were asked to grade the degraded image relative to this ground truth image as reference (device-standard, high-quality image was considered as reference image). All acquired 1764 2D image slices together with their image quality annotations are made publicly available to the research community. To the best of our knowledge, no publicly available CBCT dataset with expert clinical annotations of image quality levels, specifically developed for image quality assessment research, currently exists. Therefore, our dataset is the first publicly available CBCT image quality dataset along with clinically annotated image quality labels, establishing a valuable benchmark for the development, validation, and comparison of image quality assessment methods.

In addition to providing the dataset and annotations, we evaluated the performance of 26 IQA measures, covering both full-reference and no-reference categories and compared them against expert annotations to provide benchmark results for future IQA measures development. Furthermore, we provide an IQA measure–based rating alongside manual observer assessments that is capable of capturing subtle differences in image quality. This can be beneficial for the development of new IQA measures, enabling researchers to assess whether their methods agree with established measures in detecting and quantifying fine-grained image quality variations. The overall study design is illustrated in Figure 1.

The prepared open-access CBCT dataset together with image quality annotations, the IQA benchmark results as well as IQA measure–based rating, provides a highly valuable resource for the development and validation of future IQA measures methods. Furthermore, it can serve as a practical guideline for researchers in selecting and identifying suitable evaluation measures to assess their methods in CBCT imaging and its diverse clinical applications.

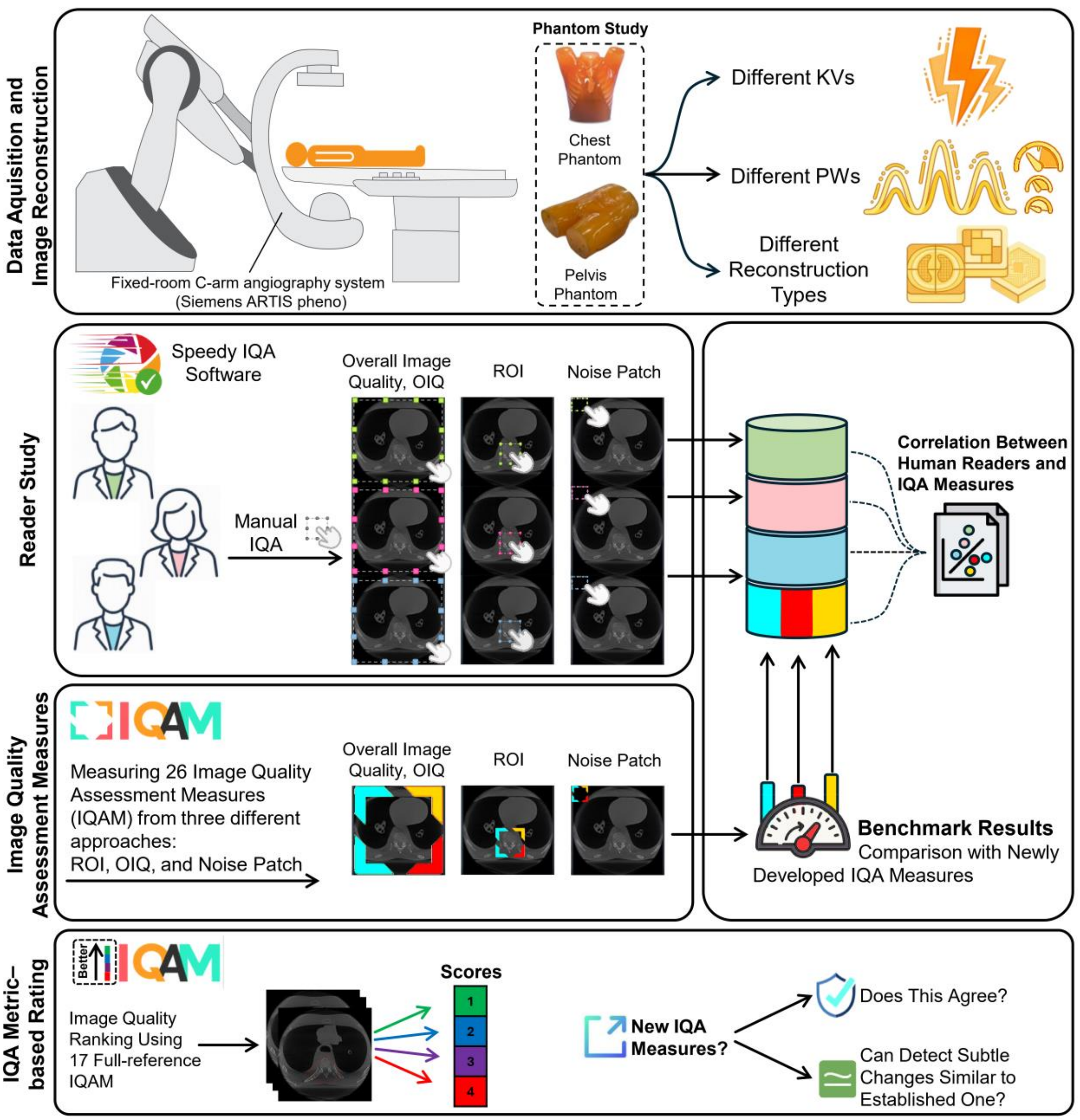

**Figure 1.** The overall proposed study design including data acquisition, reader study, image quality assessment (IQA) measures and IQA measure-based rating.

## Methods

### Phantoms

Two anatomical phantoms, a thorax and pelvis (Kyoto Kagaku), were used to generate a collection of CBCT images. The thorax phantom contains bony anatomy, including the spine, ribs, and sternum, soft tissue equivalent material, and simplified bronchial structures. The pelvis phantom contains boy anatomy, including the pelvis and hip, along with soft tissue equivalent material.

### Image Acquisition Protocols

CBCT acquisitions of the used phantoms were performed using a Siemens ARTIS pheno C-arm device (Siemens Healthcare GmbH, Erlangen, Germany), located at the University of Sydney and housed within a research-dedicated hybrid theatre.

Different acquisition parameters of the 5sDCT Body protocol (397 projections, Siemens Healthineers) were systematically varied to image the phantoms to acquire the CBCT dataset, including the kV, pulse width, and the reconstruction type (RT). Unless otherwise specified, the acquisition parameters remained unchanged from the default values for the protocol. For the generation of the dataset, the default parameters include kV of 90, pulse width of 8.0 ms and RT Normal. Images acquired using these default settings represent the device-standard, high-quality images and served as the 'reference' dataset in this study for calculating the full-reference image quality measures described in the 'Image annotation and reader study' Section. First, the kV was varied and set to the predefined values available on the system of 60, 90, 125 kV, with all other acquisition parameters remaining the same. Next, the pulse width was varied and set to the predefined values available on the system of 3.2ms, 5.0ms, 6.4ms, 8.0ms, with all other acquisition parameters remaining the same. Following acquisition, each volume was rereconstructed using the in-built system reconstructor and systematically varying the reconstruction types (RTs) including Normal, Smooth, and Very Smooth (VSmooth). As such, 36 volumes consisting of 378 slices were generated. For the pelvis dataset, we selected every third slice from slice 193 to 271, resulting in a total of 27 slices for each para-meter configuration. Similarly, for the chest dataset, we selected every third slice from slice 85 to 295, resulting in a total of 71 slices for each parameter configuration. Every third slice was selected because there were only minimal anatomical changes between consecutive slices, making this sampling strategy sufficient to capture the relevant image variations. In total, 1764 slices were used. Slice thickness and pixel spacing was set to 0.474 mm. For the Chest and Pelvis dataset, we obtained ROI image patches by cropping the full image to the pixel ranges 160-352x250-410 and 60-452x260-420, respectively, and 100x100 noise patches by cropping to 250-349x100-199. We selected the ROI region heuristically to best fit the anatomical area. The noise patch region was chosen to only contain background noise across all images. The purpose of studying the ROI region was to investigate, how the performance of the IQA measures change, when only a small portion of the image is used and background noise and unimportant regions are excluded. We also investigated the performance of IQA measures on the noise region to study, if some algorithms mainly depend on the background noise structure or the ROI. Examples of images acquired using different imaging settings are shown in Figure 2 and 3. In addition, a summary of the imgage acquisition parameters used for the 36 volumes used in this study is provided in Table 1.

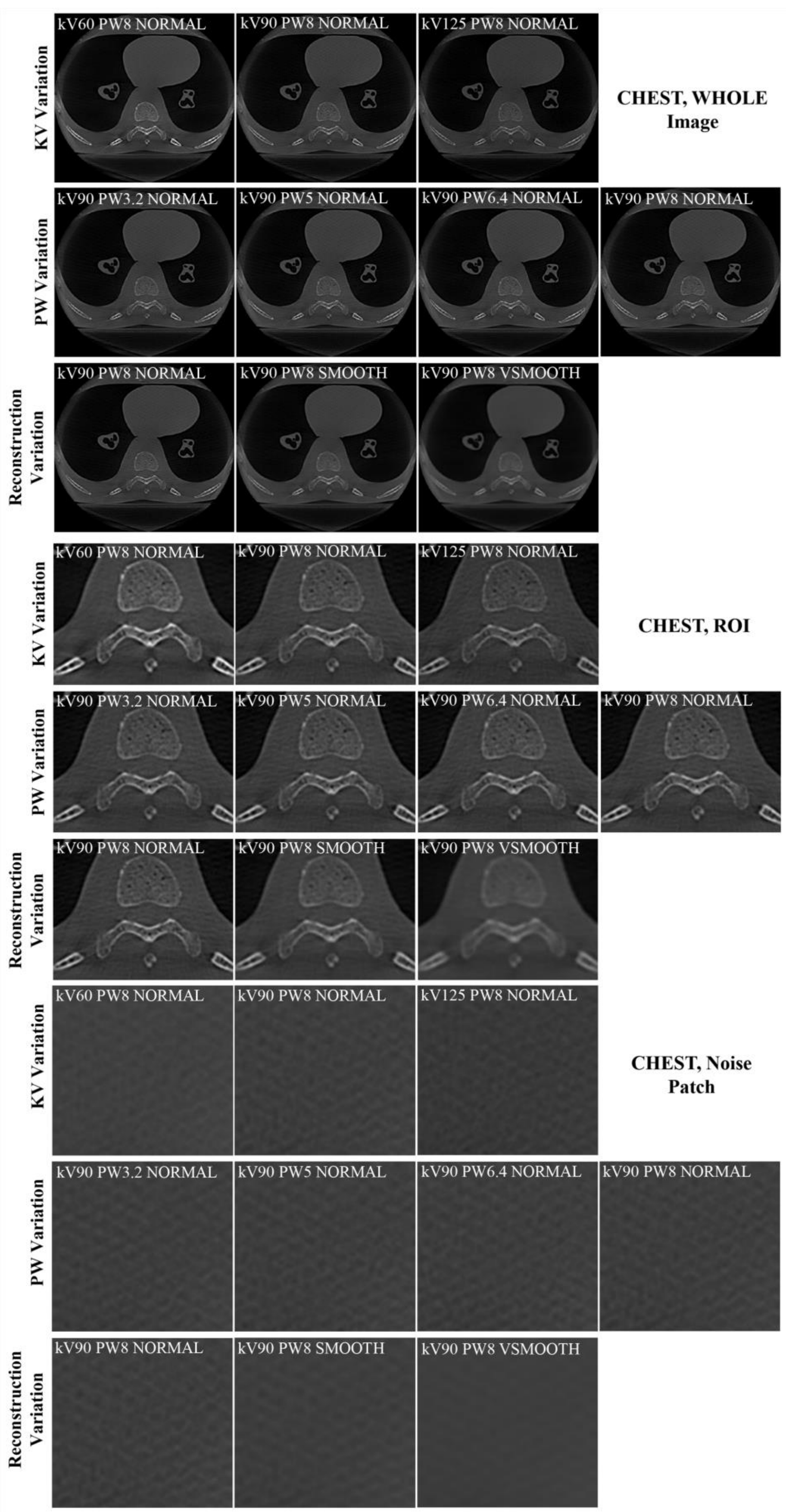

**Figure 2.** Visualization of some example used CBCT slices from thorax phantom when changing kV (60, 90, 125), pulse width (PW) (3.2, 5, 6.4, 8 ms), and reconstruction type (Norml, Smooth and very smooth (VSmooth) for overal image, Region of interest (ROI) and noise patch.

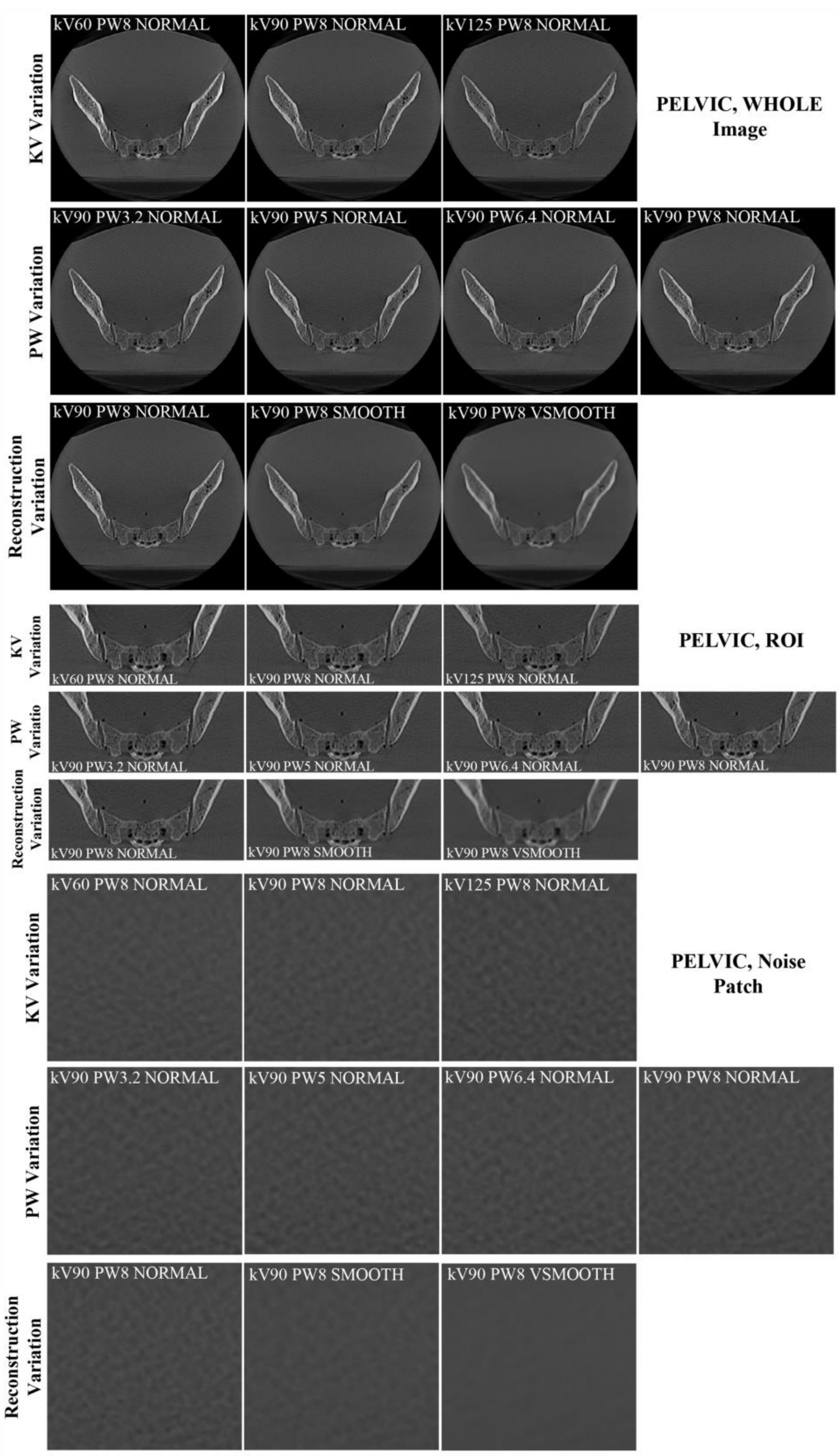

**Figure 3.** Visualization of some example used CBCT slices from pelvis phantom when changing kV (60, 90, 125), pulse width (PW) (3.2, 5, 6.4, 8 ms), and reconstruction type (Norml, Smooth and very smooth (VSmooth) for overal image, Region of interest (ROI) and noise patch.

| Dataset | kV | Pulse width | Number of slices | Reconstruction type | Type of image |
|---|---|---|---|---|---|
| **Chest** | 60 | 8 | 71 | Nomal | Degrded |
| **Chest** | 60 | 8 | 71 | Smooth | Degrded |
| **Chest** | 60 | 8 | 71 | Very Smooth | Degrded |
| **Chest** | 90 | 8 | 71 | Nomal | Reference |
| **Chest** | 90 | 8 | 71 | Smooth | Degrded |
| **Chest** | 90 | 8 | 71 | Very Smooth | Degrded |
| **Chest** | 125 | 8 | 71 | Nomal | Degrded |
| **Chest** | 125 | 8 | 71 | Smooth | Degrded |
| **Chest** | 125 | 8 | 71 | Very Smooth | Degrded |
| **Chest** | 90 | 3,2 | 71 | Nomal | Degrded |
| **Chest** | 90 | 3,2 | 71 | Smooth | Degrded |
| **Chest** | 90 | 3,2 | 71 | Very Smooth | Degrded |
| **Chest** | 90 | 5 | 71 | Nomal | Degrded |
| **Chest** | 90 | 5 | 71 | Smooth | Degrded |
| **Chest** | 90 | 5 | 71 | Very Smooth | Degrded |
| **Chest** | 90 | 6,4 | 71 | Nomal | Degrded |
| **Chest** | 90 | 6,4 | 71 | Smooth | Degrded |
| **Chest** | 90 | 6,4 | 71 | Very Smooth | Degrded |
| **Pelvis** | 60 | 8 | 27 | Nomal | Degrded |
| **Pelvis** | 60 | 8 | 27 | Smooth | Degrded |
| **Pelvis** | 60 | 8 | 27 | Very Smooth | Degrded |
| **Pelvis** | 90 | 8 | 27 | Nomal | Reference |
| **Pelvis** | 90 | 8 | 27 | Smooth | Degrded |
| **Pelvis** | 90 | 8 | 27 | Very Smooth | Degrded |
| **Pelvis** | 125 | 8 | 27 | Nomal | Degrded |
| **Pelvis** | 125 | 8 | 27 | Smooth | Degrded |
| **Pelvis** | 125 | 8 | 27 | Very Smooth | Degrded |
| **Pelvis** | 90 | 3,2 | 27 | Nomal | Degrded |
| **Pelvis** | 90 | 3,2 | 27 | Smooth | Degrded |
| **Pelvis** | 90 | 3,2 | 27 | Very Smooth | Degrded |
| **Pelvis** | 90 | 5 | 27 | Nomal | Degrded |
| **Pelvis** | 90 | 5 | 27 | Smooth | Degrded |
| **Pelvis** | 90 | 5 | 27 | Very Smooth | Degrded |
| **Pelvis** | 90 | 6,4 | 27 | Nomal | Degrded |
| **Pelvis** | 90 | 6,4 | 27 | Smooth | Degrded |
| **Pelvis** | 90 | 6,4 | 27 | Very Smooth | Degrded |

**Table 1.** Summary of the acquisition scheme for the 36 volumes used.

### Image pre-processing and normalization

The used slices were normalized, by first computing the minimal and maximal pixel values over all reference slices of the volume. Then, for each selected slice not to be used as reference data, we clipped pixel values larger than the maximal pixel value in the reference slices to the reference slices maximal value. Then we subtracted from every slice the minimal pixel value over all reference slices and divided by the maximum reference pixel value minus the minimal reference pixel value. Let $r_{min}$ and $r_{max}$ be the minimal and maximal pixel values over all reference slices of a volume. Then, to compute the normalized image $I$, we apply to each slice $S$ the normalization $I_{x,y} = \frac{min(S_{x,y}, r_{max}) - r_{min}}{r_{max} - r_{min}}$.

### Image annotation and reader study

**Annotation software and configuration**

For expert-based image quality assessment, the open-source software Speedy IQA [25] was used. This software enables paired comparison of test- and reference-images and supports structured quality annotation workflows. Prior to annotation, the labelling structure was configured within the software. The main task category "Image-guided therapy" and the subcategory "Region of Interest (ROI)" were defined.

The image dataset consisted of:

- Test images (degraded images to be evaluated) and
- Reference images (reference images used for comparison)

During each annotation session, a test image and its corresponding reference image were displayed simultaneously. The image presentation order was randomized. Images were rated using a 4-point Likert scale, from rank 1 = very poor to rank 4 = very good.

Two assessment dimensions were evaluated:

1. Overall image quality (OIQ) (entire image, see Figure 4)
2. Quality within a predefined region of interest (ROI) (defined region, see Figure 5)

The software supports zooming and panning to allow detailed inspection of the ROI. In advance, detailed information (Figure 5, Left) was sent to the annotators, and which area should be assessed as ROI in detail (see Figure 5, Middle and Right).

**Annotation procedure and readers**

Three independent expert raters performed the image assessments. For each case, raters were shown the reference (ground truth) image simultaneously with the corresponding degraded image and were asked to assess the quality of the degraded image relative to the reference. Each rater performed the two assessment tasks (overall quality and ROI-based quality) by image.

The image assessment was conducted by three specialists with extensive expertise in oncologic imaging and image quality evaluation:

- Rater 1 is a board-certified radiologist with more than 15 years of experience in oncologic imaging. Her expertise includes advanced imaging techniques, quantitative image analysis, radiomics, and clinical as well as research-based image quality evaluation
- Rater 2 is a radiologist and Research Fellow at the European School of Radiology (ESOR), with substantial clinical and research experience in oncological imaging and dedicated expertise in image quality evaluation
- Rater 3 is a medical physicist with more than 10 years of experience in radiation oncology and image-guided radiotherapy in both clinical and research settings. His expertise includes image quality assurance of X-ray–based imaging systems and cone-beam CT (CBCT)-guided therapies

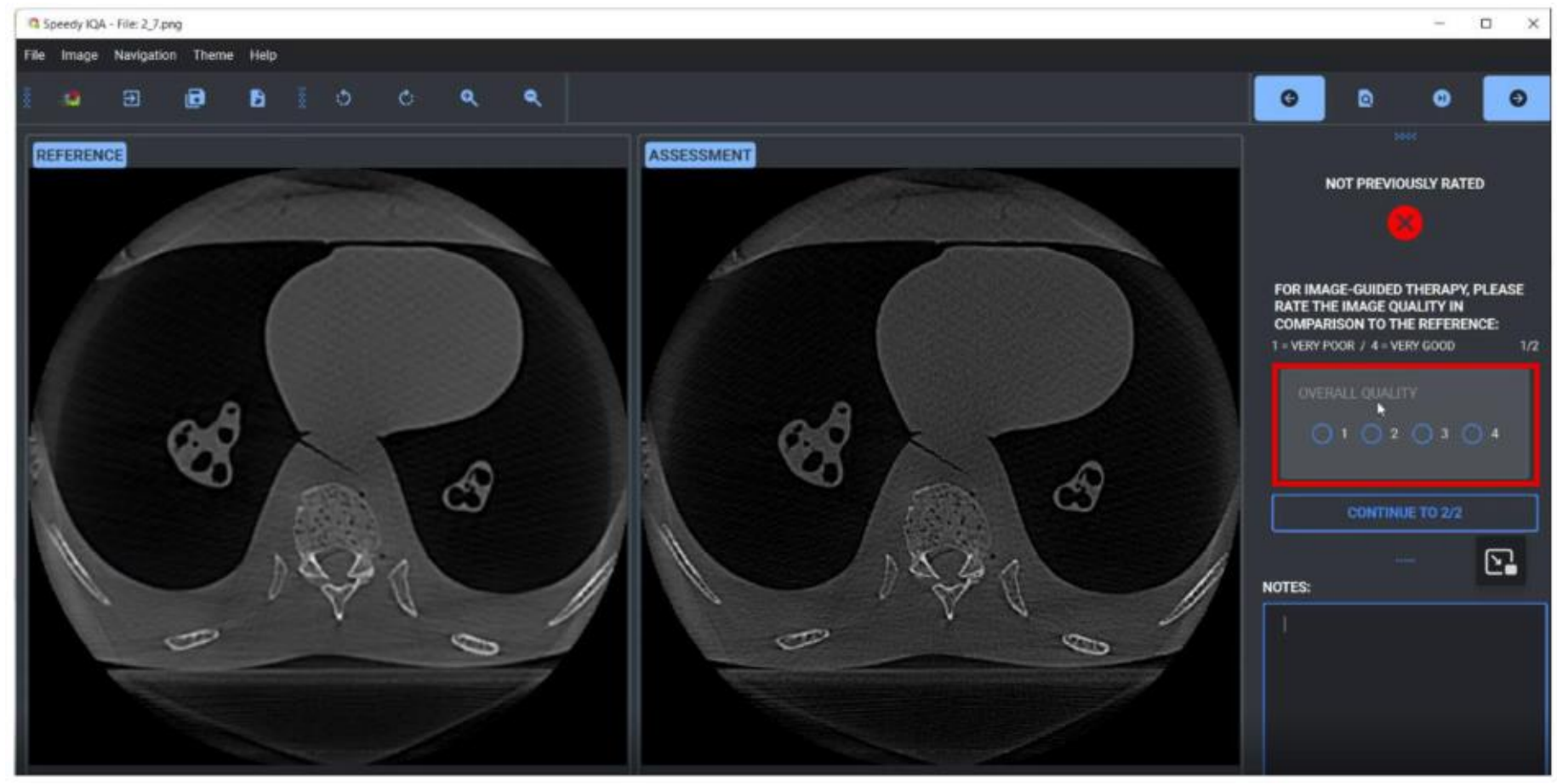

**Figure 4.** A screenshot from open-source software Speedy IQA for overall image quality (OIQ) assessment.

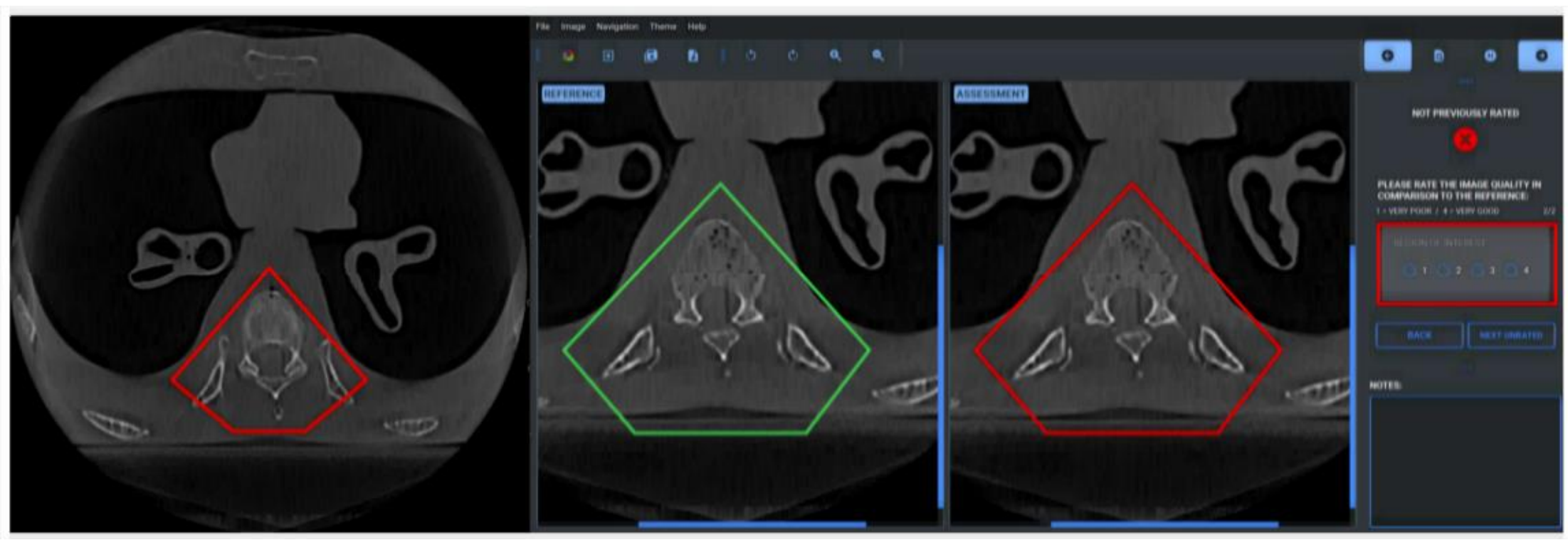

**Figure 5**. Visualization of image together with the defined region of interest (ROI) (Left). A screenshot from open-source software Speedy IQA for ROI assessment (Middle, Right).

**Correlation between different annotators' ratings**

To assess the similarity between annotators' ratings, we employed the Spearman Rank Correlation Coefficient (SRCC), which measures the ordinal association (ranking agreement) between different annotators' scores for both whole-image and ROI image assessment cases. To account for the different scoring between multiple graders, we apply the z-score to the raw rating data of each grader and afterwards compute the mean score similar to [26].
The correlation was computed between annotators for three cases: OIQ vs. OIQ, ROI vs. ROI, and ROI vs. OIQ (Table 2). In the case of OIQ vs. OIQ, the correlation was consistently high (above 0.80). While for annotation ratings of ROI vs. ROI and ROI vs. OIQ, annotator 3 showed a lower correlation (Table 2).

| | **OIQ-1** | **OIQ -2** | **OIQ -3** |
|---|---|---|---|
| OIQ -1 | - | 0.88 | 0.81 |
| OIQ -2 | - | - | 0.80 |
| OIQ -3 | - | - | - |
| ROI-1 | 0.93 | 0.88 | 0.81 |
| ROI-2 | 0.81 | 0.81 | 0.74 |
| ROI-3 | 0.70 | 0.69 | 0.63 |
| | **ROI-1** | **ROI-2** | **ROI-3** |
| ROI-1 | - | 0.82 | 0.71 |

| ROI-2 | - | - | 0.66 |
|---|---|---|---|
| ROI-3 | - | - | - |

**Table 2.** Spearman Rank Correlation Coefficient (SRCC) between different annotators' scores for both overall image quality (OIQ) and region of interest (ROI) image assessment cases.

## Evaluation of annotators' rating sensitivity to image acquisition variations

For both OIQ and ROI, changes due to different kV and pulse width values were visually not very apparent, whereas differences across RTs were visually clearly noticeable (Figures 2, 3). Consistent with this, observers' ratings were very similar (mainly one rank score was assigned) when annotating images with varying kV and pulse width and fixed kernel (Figure 6, rows B, C, D), suggesting that such changes in image quality were not easily detectable, confirmed by visual inspection. The majority of the images for the RT Normal (row B), Smooth (row C) and VSmooth (row D) with varied KV or PW are rated 4, 3 to 4, and 1, respectively, which implies that the KV and PW variation in this setting is not discernible for the annotators for the rated task.

In contrast, variations across RTs were evident visually (Figures 2, 3) and easily captured by observers in their rating i.e., observers rating were different (different rates were assigned i.e., images with varied RT and fixed KV or PW are rated from 1 to 4, covering the whole range) (Figure 6, row A).

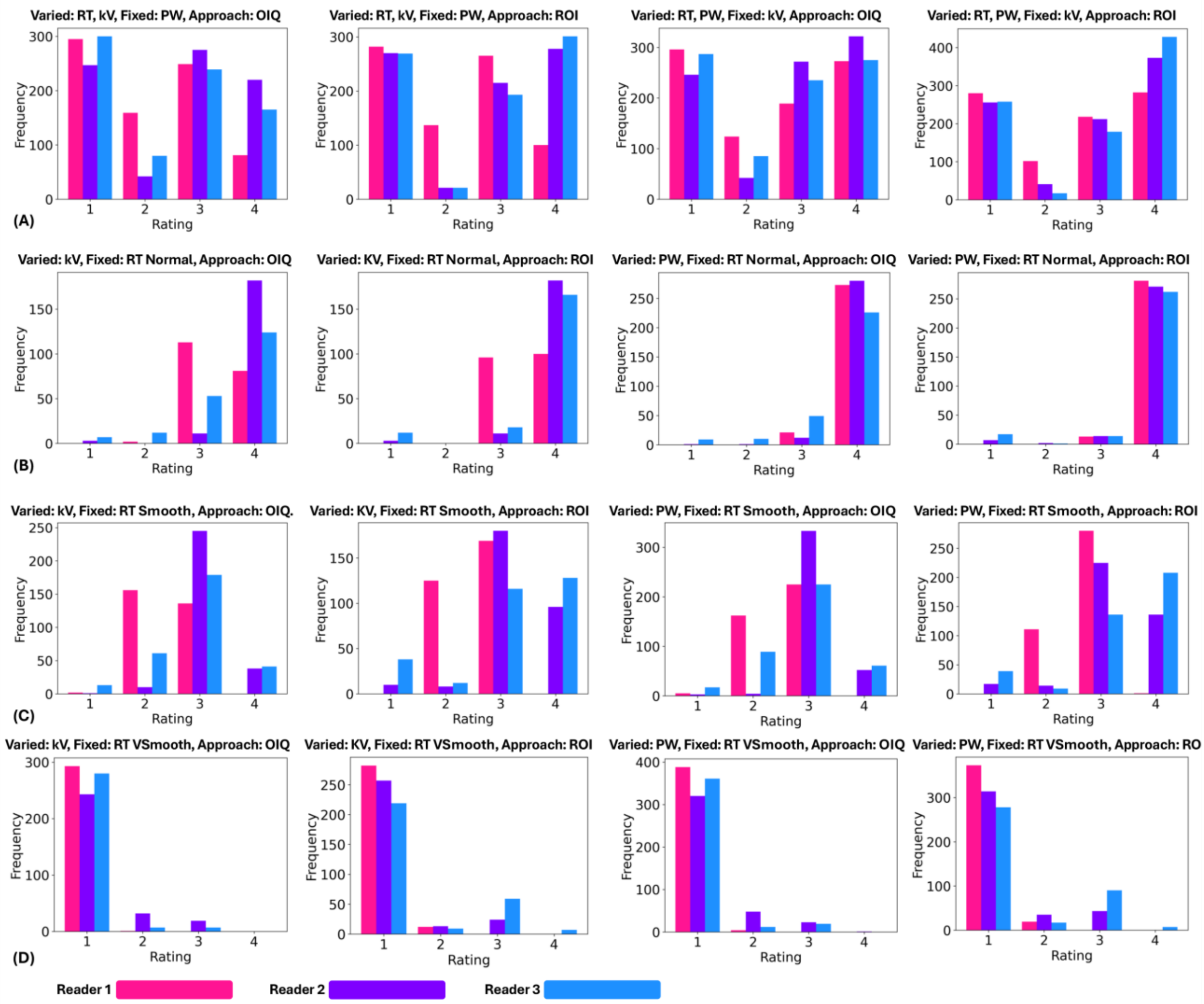

**Figure 6.** Distribution of image quality ratings assigned by the three expert readers (ratings 1–4) across different imaging scenarios. (A) Images acquired with varying reconstruction types (RT) while keeping either the tube voltage (kV) or pulse width (PW) constant. (B–D) Images acquired with varying kV and pulse width while maintaining a fixed reconstruction type: Normal (B), Smooth (C), and Very Smooth (D). Ratings are shown separately for overall image quality (OIQ) and region-of-interest (ROI) image quality assessments.

## Image quality assessment (IQA)

### Quantitative IQA analysis of different existing measures

In order to make a benchmark on the performance of different IQA measures for CBCT imaging, we compared the performance of 26 commonly used as well as task-based IQA measures (27 IQA measure implementations) covering both full-reference and no-reference-based IQA measures categories, against expert annotations/manual ratings. For full-reference IQA measure calculation, the CBCT image acquired with default parameters (kV of 90, pulse width of 8.0 ms and RT Normal, Section 'Image Acquisition Protocols') was used as the reference image, as it represents the highest image quality. The IQA measures used include structural similarity index measure (SSIM) [27], multi-scale SSIM (MS-SSIM) [28], information content weighted SSIM (IW-SSIM) [29], Complex-Wavelet Structural Similarity Index (CW-SSIM) [30], peak signal-to-noise ratio (PSNR) [31], root-mean-square error (RMSE) [32], deep image structure and texture similarity (DISTS) [33], DCT subband similarity (DSS) [34], feature similarity index (FSIM) [35], gradient magnitude similarity deviation (GMSD) [36], Haar wavelet-based perceptual similarity index (HaarPSI) [37], HaarPSImed [38], learned perceptual image patch similarity (LPIPS) [39], mean deviation similarity (MDSI) [40], visual information fidelity (VIF) [41], visual saliency-induced index (VSI) [42], Average Mean Brightness Error (AMBE) [43], natural image quality evaluator (NIQE) [44], Signal-to-noise ratio (SNR) [45], from patches to pictures (PaQ-2-PIQ) [46], Perception based Image Quality Evaluator (PIQE) [47], CHILL@UK [48], FeatureNet [48], Gabybaldeon [48], RPIAXIS [48], Team Epoch [48].

For this purpose, we computed the SRCC (using the same methods described in Section 'Correlation Between Different Annotators' Ratings') between the IQA measures and the ratings provided by the graders. The correlations, computed for both OIQ and ROI, computed over all dataset (both phantom types, all kV and pulse width variations and RTs), are presented in Table 3. The empty entries in Tables 3–6 indicate that some image quality assessment algorithms could not be computed because they require a minimum image size that was not met by a subset of the images in our dataset.
According to our results, the full-reference-based IQA measures with highest correlations with expert annotations for both whole-image and ROI evaluations (Table 3) were LPIPS-python, DISTS-python/matlab, FSIM-matlab, VSI-matlab, VIF-python, DSS-matlab, and all Haar-based measures. Among the non-reference-based methods, CHILLUK-python, TeamEpock-python, RPIAXIS-python, PIQE-matlab and FeatureNet-python showed highest correlation.

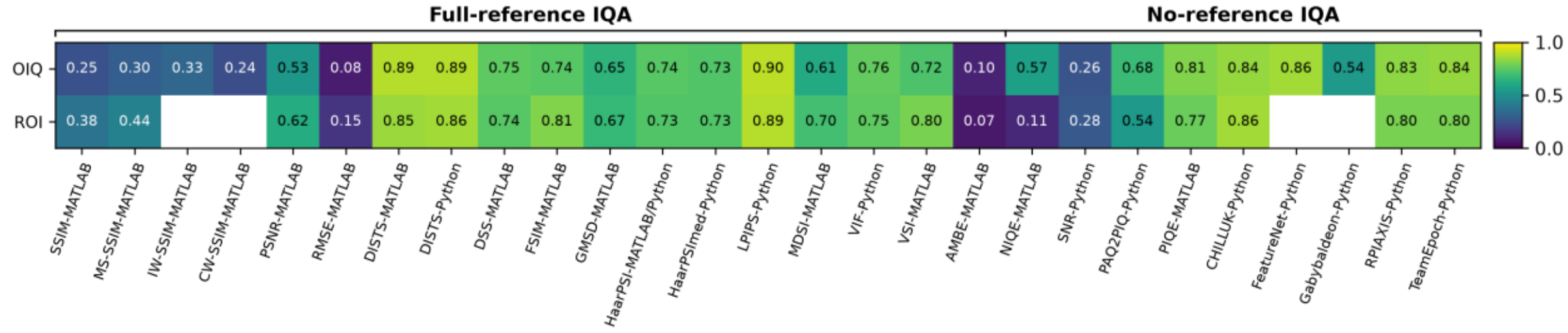

| | Full-reference IQA | | | | | | | | | | | | | | | | |
|---|---|---|---|---|---|---|---|---|---|---|---|---|---|---|---|---|---|
| | SSIM-MATLAB | MS-SSIM-MATLAB | IW-SSIM-MATLAB | CW-SSIM-MATLAB | PSNR-MATLAB | RMSE-MATLAB | DISTS-MATLAB | DISTS-Python | DSS-MATLAB | FSIM-MATLAB | GMSD-MATLAB | HaarPSI-MATLAB/Python | HaarPSImed-Python | LPIPS-Python | MDSI-MATLAB | VIF-Python | VSI-MATLAB |
| OIQ | 0.25 | 0.30 | 0.33 | 0.24 | 0.53 | 0.08 | 0.89 | 0.89 | 0.75 | 0.74 | 0.65 | 0.74 | 0.73 | 0.90 | 0.61 | 0.76 | 0.72 |
| ROI | 0.38 | 0.44 | | | 0.62 | 0.15 | 0.85 | 0.86 | 0.74 | 0.81 | 0.67 | 0.73 | 0.73 | 0.89 | 0.70 | 0.75 | 0.80 |

| | No-reference IQA | | | | | | | | | |
|---|---|---|---|---|---|---|---|---|---|---|
| | AMBE-MATLAB | NIQE-MATLAB | SNR-Python | PAQ2PIQ-Python | PIQE-MATLAB | CHILLUK-Python | FeatureNet-Python | Gabybaldeon-Python | RPIAXIS-Python | TeamEpoch-Python |
| OIQ | 0.10 | 0.57 | 0.26 | 0.68 | 0.81 | 0.84 | 0.86 | 0.54 | 0.83 | 0.84 |
| ROI | 0.07 | 0.11 | 0.28 | 0.54 | 0.77 | 0.86 | | | 0.80 | 0.80 |

**Table 3.** Spearman's rank correlation coefficient (SRCC) values, computed across all imaging datasets, phantoms, and image acquisition settings, between each of the 27 calculated image quality assessment (IQA) measures and the expert ratings for both overall image quality (OIQ) and region-of-interest (ROI) image quality analyses.

## Performance of IQA measures in presence of background noise

We further analyzed the results to evaluate the performance of these measures in relation to noise. A patch from the background noise was selected in all cases (see examples in Figures 2 and 3), and the IQA measures were calculated on this noise patch (Table 4). Although a large drop in performance was expected when applying the measures to pure noise patches, we observed that some IQA measures with high correlation performed similarly across the whole image, ROI, and pure noise patches; this includes measures LPIPS-python, DISTS-python/matlab, TeamEpock-python, and RPIAXIS-python. This indicates that these measures are primarily responding to background noise rather than capturing structural information.

We also observed that noise in the reconstruction type Normal was more pronounced visually compared to the other two types (Smooth and Very Smooth) (Figures 2 and 3). Therefore, as a second experiments to assess the influence of the noise, we excluded the images with reconstruction type Normal from the experiment and repeated the calculations (Table 4, rows 'OIQ Normal Excluded' and 'ROI Normal Excluded') to evaluate the performance of the IQ measures when this strong noise was removed from the dataset. A considerable drop in performance for both OIQ and ROI were observed for the IQA measures DISTS-python/matlab, LPIPS-python, and RPIAXIS, further confirming higher sensitivity of these measures to background noise. Based on these results, we conclude that the remaining IQA measures including PIQE-matlab (non-reference) and CHILLUK-python as well as FSIM-matlab, VIF-python, Haar-based measures, DSS-matlab, and VSI-matlab, were the best-performing measures, considering both their high correlation values to manual rating and their robustness against noise.

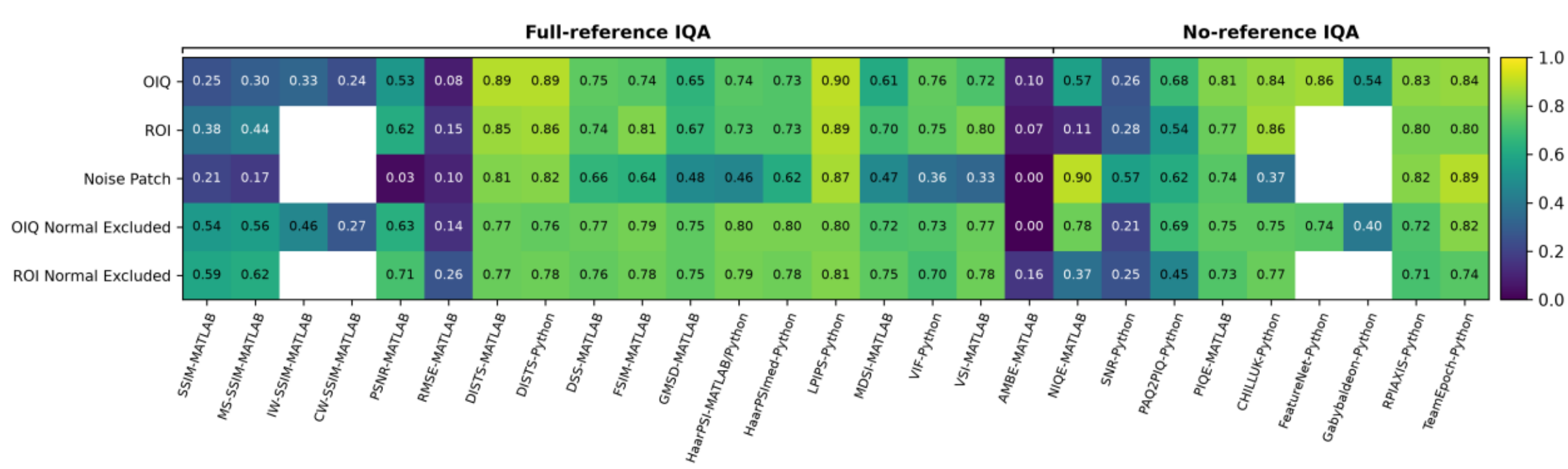

| | Full-reference IQA | | | | | | | | | | | | | | | | |
|---|---|---|---|---|---|---|---|---|---|---|---|---|---|---|---|---|---|
| | SSIM-MATLAB | MS-SSIM-MATLAB | IW-SSIM-MATLAB | CW-SSIM-MATLAB | PSNR-MATLAB | RMSE-MATLAB | DISTS-MATLAB | DISTS-Python | DSS-MATLAB | FSIM-MATLAB | GMSD-MATLAB | HaarPSI-MATLAB/Python | HaarPSImed-Python | LPIPS-Python | MDSI-MATLAB | VIF-Python | VSI-MATLAB |
| OIQ | 0.25 | 0.30 | 0.33 | 0.24 | 0.53 | 0.08 | 0.89 | 0.89 | 0.75 | 0.74 | 0.65 | 0.74 | 0.73 | 0.90 | 0.61 | 0.76 | 0.72 |
| ROI | 0.38 | 0.44 | | | 0.62 | 0.15 | 0.85 | 0.86 | 0.74 | 0.81 | 0.67 | 0.73 | 0.73 | 0.89 | 0.70 | 0.75 | 0.80 |
| Noise Patch | 0.21 | 0.17 | | | 0.03 | 0.10 | 0.81 | 0.82 | 0.66 | 0.64 | 0.48 | 0.46 | 0.62 | 0.87 | 0.47 | 0.36 | 0.33 |
| OIQ Normal Excluded | 0.54 | 0.56 | 0.46 | 0.27 | 0.63 | 0.14 | 0.77 | 0.76 | 0.77 | 0.79 | 0.75 | 0.80 | 0.80 | 0.80 | 0.72 | 0.73 | 0.77 |
| ROI Normal Excluded | 0.59 | 0.62 | | | 0.71 | 0.26 | 0.77 | 0.78 | 0.76 | 0.78 | 0.75 | 0.79 | 0.78 | 0.81 | 0.75 | 0.70 | 0.78 |

| | No-reference IQA | | | | | | | | | |
|---|---|---|---|---|---|---|---|---|---|---|
| | AMBE-MATLAB | NIQE-MATLAB | SNR-Python | PAQ2PIQ-Python | PIQE-MATLAB | CHILLUK-Python | FeatureNet-Python | Gabybaldeon-Python | RPIAXIS-Python | TeamEpoch-Python |
| OIQ | 0.10 | 0.57 | 0.26 | 0.68 | 0.81 | 0.84 | 0.86 | 0.54 | 0.83 | 0.84 |
| ROI | 0.07 | 0.11 | 0.28 | 0.54 | 0.77 | 0.86 | | | 0.80 | 0.80 |
| Noise Patch | 0.00 | 0.90 | 0.57 | 0.62 | 0.74 | 0.37 | | | 0.82 | 0.89 |
| OIQ Normal Excluded | 0.00 | 0.78 | 0.21 | 0.69 | 0.75 | 0.75 | 0.74 | 0.40 | 0.72 | 0.82 |
| ROI Normal Excluded | 0.16 | 0.37 | 0.25 | 0.45 | 0.73 | 0.77 | | | 0.71 | 0.74 |

**Table 4.** Spearman's rank correlation coefficient (SRCC) values, computed across the combined imaging datasets encompassing all phantoms and image acquisition settings, between each of the 27 calculated image quality assessment (IQA) measures and the expert ratings for overall image quality (OIQ), region-of-interest (ROI) image quality, noise patch analysis, OIQ excluding images acquired

with the Normal reconstruction type (OIQ Normal Excluded), and ROI excluding images acquired with the Normal reconstruction type (ROI Normal Excluded).

## Detailed evaluation of the correlation between annotators and IQA measures with respect to the image acquisition variations

We performed some experiments to evaluate the sensitivity of different IQA measures to acquisition and RT variations. To assess the correlation values when changing kV and pulse width, we restricted experiments to a single reconstruction type at a time (to not include variation in reconstruction types) while allowing variation in kV and pulse width. The correlation results between annotators and IQA measures for the OIQ, ROI and noise are shown in Table 5. Our results showed that when experiments were limited to a single reconstruction type, the correlation between annotators and IQA measures dropped dramatically. This indicates that the high correlation between manual ratings and IQA measures reported in Table 4 is driven by reconstruction type changes rather than by kV or pulse width. To further verify this, we analyzed a subset with fixed kV and pulse width (PW) (PW8-KV90=pulse width 8 and kV 90) for OQI, ROI, and noise to remove the effect of variation of kV and pulse width while keeping purely the effect the variation of reconstruction type) (Table 6). For this subset, we observed very high correlations, confirming that the strong correlations mainly stem from reconstruction type variations.

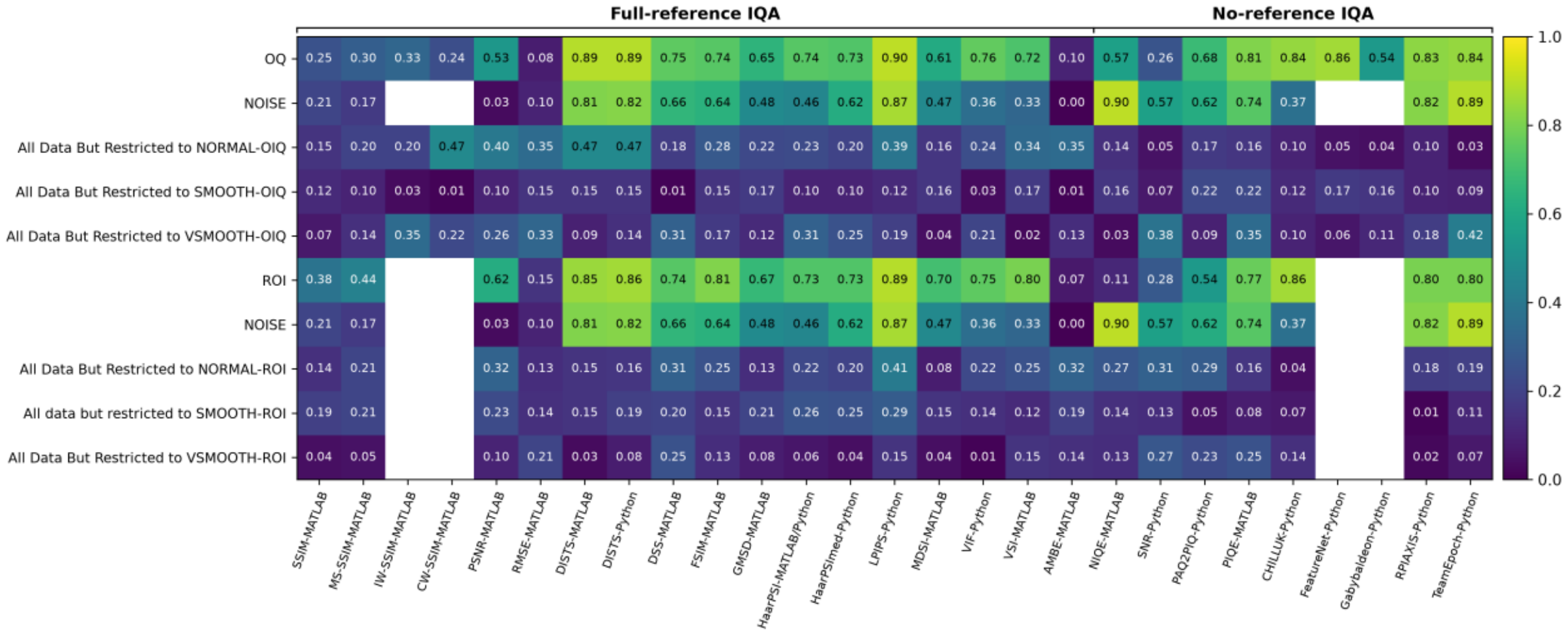

Full-reference IQA

| | SSIM-MATLAB | MS-SSIM-MATLAB | IW-SSIM-MATLAB | CW-SSIM-MATLAB | PSNR-MATLAB | RMSE-MATLAB | DISTS-MATLAB | DISTS-Python | DSS-MATLAB | FSIM-MATLAB | GMSD-MATLAB | HaarPSI-MATLAB/Python | HaarPSImed-Python | LPIPS-Python | MDSI-MATLAB | VIF-Python | VSI-MATLAB | AMBE-MATLAB |
|---|---|---|---|---|---|---|---|---|---|---|---|---|---|---|---|---|---|---|
| OQ | 0.25 | 0.30 | 0.33 | 0.24 | 0.53 | 0.08 | 0.89 | 0.89 | 0.75 | 0.74 | 0.65 | 0.74 | 0.73 | 0.90 | 0.61 | 0.76 | 0.72 | 0.10 |
| NOISE | 0.21 | 0.17 | | | 0.03 | 0.10 | 0.81 | 0.82 | 0.66 | 0.64 | 0.48 | 0.46 | 0.62 | 0.87 | 0.47 | 0.36 | 0.33 | 0.00 |
| All Data But Restricted to NORMAL-OIQ | 0.15 | 0.20 | 0.20 | 0.47 | 0.40 | 0.35 | 0.47 | 0.47 | 0.18 | 0.28 | 0.22 | 0.23 | 0.20 | 0.39 | 0.16 | 0.24 | 0.34 | 0.35 |
| All Data But Restricted to SMOOTH-OIQ | 0.12 | 0.10 | 0.03 | 0.01 | 0.10 | 0.15 | 0.15 | 0.15 | 0.01 | 0.15 | 0.17 | 0.10 | 0.10 | 0.12 | 0.16 | 0.03 | 0.17 | 0.01 |
| All Data But Restricted to VSMOOTH-OIQ | 0.07 | 0.14 | 0.35 | 0.22 | 0.26 | 0.33 | 0.09 | 0.14 | 0.31 | 0.17 | 0.12 | 0.31 | 0.25 | 0.19 | 0.04 | 0.21 | 0.02 | 0.13 |
| ROI | 0.38 | 0.44 | | | 0.62 | 0.15 | 0.85 | 0.86 | 0.74 | 0.81 | 0.67 | 0.73 | 0.73 | 0.89 | 0.70 | 0.75 | 0.80 | 0.07 |
| NOISE | 0.21 | 0.17 | | | 0.03 | 0.10 | 0.81 | 0.82 | 0.66 | 0.64 | 0.48 | 0.46 | 0.62 | 0.87 | 0.47 | 0.36 | 0.33 | 0.00 |
| All Data But Restricted to NORMAL-ROI | 0.14 | 0.21 | | | 0.32 | 0.13 | 0.15 | 0.16 | 0.31 | 0.25 | 0.13 | 0.22 | 0.20 | 0.41 | 0.08 | 0.22 | 0.25 | 0.32 |
| All data but restricted to SMOOTH-ROI | 0.19 | 0.21 | | | 0.23 | 0.14 | 0.15 | 0.19 | 0.20 | 0.15 | 0.21 | 0.26 | 0.25 | 0.29 | 0.15 | 0.14 | 0.12 | 0.19 |
| All Data But Restricted to VSMOOTH-ROI | 0.04 | 0.05 | | | 0.10 | 0.21 | 0.03 | 0.08 | 0.25 | 0.13 | 0.08 | 0.06 | 0.04 | 0.15 | 0.04 | 0.01 | 0.15 | 0.14 |

No-reference IQA

| | NIQE-MATLAB | SNR-Python | PAQ2PIQ-Python | PIQE-MATLAB | CHILLUK-Python | FeatureNet-Python | Gabybaldeon-Python | RPIAXIS-Python | TeamEpoch-Python |
|---|---|---|---|---|---|---|---|---|---|
| OQ | 0.57 | 0.26 | 0.68 | 0.81 | 0.84 | 0.86 | 0.54 | 0.83 | 0.84 |
| NOISE | 0.90 | 0.57 | 0.62 | 0.74 | 0.37 | | | 0.82 | 0.89 |
| All Data But Restricted to NORMAL-OIQ | 0.14 | 0.05 | 0.17 | 0.16 | 0.10 | 0.05 | 0.04 | 0.10 | 0.03 |
| All Data But Restricted to SMOOTH-OIQ | 0.16 | 0.07 | 0.22 | 0.22 | 0.12 | 0.17 | 0.16 | 0.10 | 0.09 |
| All Data But Restricted to VSMOOTH-OIQ | 0.03 | 0.38 | 0.09 | 0.35 | 0.10 | 0.06 | 0.11 | 0.18 | 0.42 |
| ROI | 0.11 | 0.28 | 0.54 | 0.77 | 0.86 | | | 0.80 | 0.80 |
| NOISE | 0.90 | 0.57 | 0.62 | 0.74 | 0.37 | | | 0.82 | 0.89 |
| All Data But Restricted to NORMAL-ROI | 0.27 | 0.31 | 0.29 | 0.16 | 0.04 | | | 0.18 | 0.19 |
| All data but restricted to SMOOTH-ROI | 0.14 | 0.13 | 0.05 | 0.08 | 0.07 | | | 0.01 | 0.11 |
| All Data But Restricted to VSMOOTH-ROI | 0.13 | 0.27 | 0.23 | 0.25 | 0.14 | | | 0.02 | 0.07 |

**Table 5.** Spearman's rank correlation coefficient (SRCC) values, computed across all imaging datasets encompassing all phantoms and image acquisition settings, between each of the 27 calculated image quality assessment (IQA) measures and the expert ratings for overall image quality (OIQ), region-of-interest (ROI) image quality, and noise patch analysis. In addition, SRCC values are reported for OIQ and ROI analyses performed separately for each reconstruction type (Normal, Smooth, and Very Smooth), with the dataset restricted to images acquired using the respective reconstruction type.

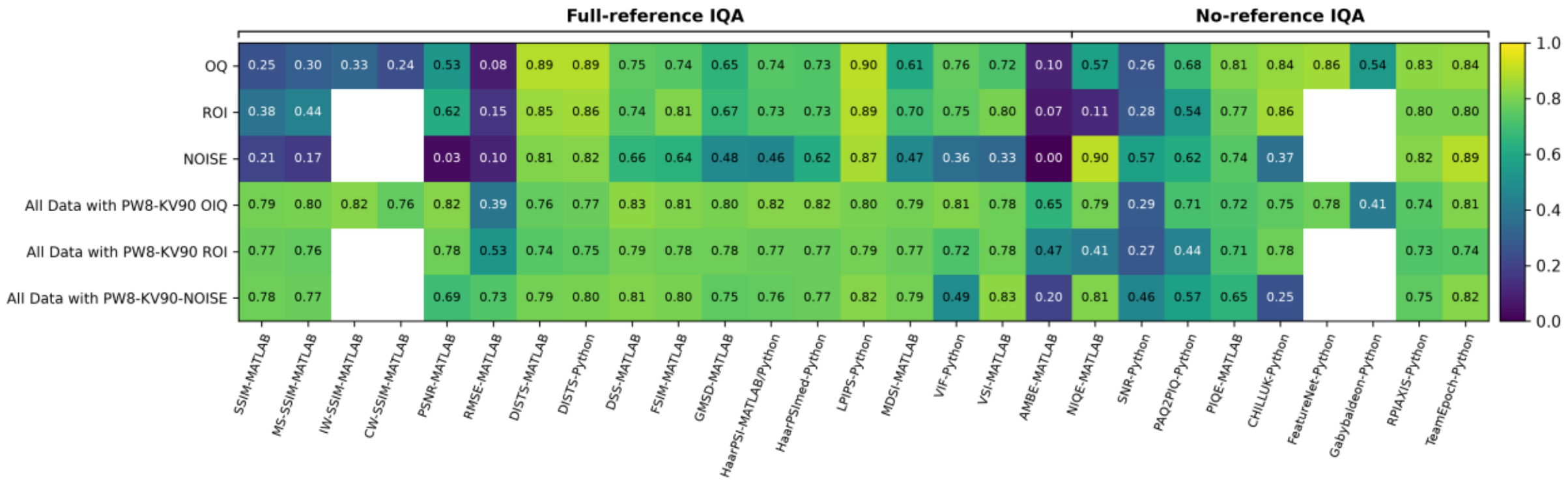

Full-reference IQA

| | SSIM-MATLAB | MS-SSIM-MATLAB | IW-SSIM-MATLAB | CW-SSIM-MATLAB | PSNR-MATLAB | RMSE-MATLAB | DISTS-MATLAB | DISTS-Python | DSS-MATLAB |
|---|---|---|---|---|---|---|---|---|---|
| OQ | 0.25 | 0.30 | 0.33 | 0.24 | 0.53 | 0.08 | 0.89 | 0.89 | 0.75 |
| ROI | 0.38 | 0.44 | | | 0.62 | 0.15 | 0.85 | 0.86 | 0.74 |
| NOISE | 0.21 | 0.17 | | | 0.03 | 0.10 | 0.81 | 0.82 | 0.66 |
| All Data with PW8-KV90 OIQ | 0.79 | 0.80 | 0.82 | 0.76 | 0.82 | 0.39 | 0.76 | 0.77 | 0.83 |
| All Data with PW8-KV90 ROI | 0.77 | 0.76 | | | 0.78 | 0.53 | 0.74 | 0.75 | 0.79 |
| All Data with PW8-KV90-NOISE | 0.78 | 0.77 | | | 0.69 | 0.73 | 0.79 | 0.80 | 0.81 |

| | FSIM-MATLAB | GMSD-MATLAB | HaarPSI-MATLAB/Python | HaarPSImed-Python | LPIPS-Python | MDSI-MATLAB | VIF-Python | VSI-MATLAB | AMBE-MATLAB |
|---|---|---|---|---|---|---|---|---|---|
| OQ | 0.74 | 0.65 | 0.74 | 0.73 | 0.90 | 0.61 | 0.76 | 0.72 | 0.10 |
| ROI | 0.81 | 0.67 | 0.73 | 0.73 | 0.89 | 0.70 | 0.75 | 0.80 | 0.07 |
| NOISE | 0.64 | 0.48 | 0.46 | 0.62 | 0.87 | 0.47 | 0.36 | 0.33 | 0.00 |
| All Data with PW8-KV90 OIQ | 0.81 | 0.80 | 0.82 | 0.82 | 0.80 | 0.79 | 0.81 | 0.78 | 0.65 |
| All Data with PW8-KV90 ROI | 0.78 | 0.78 | 0.77 | 0.77 | 0.79 | 0.77 | 0.72 | 0.78 | 0.47 |
| All Data with PW8-KV90-NOISE | 0.80 | 0.75 | 0.76 | 0.77 | 0.82 | 0.79 | 0.49 | 0.83 | 0.20 |

No-reference IQA

| | NIQE-MATLAB | SNR-Python | PAQ2PIQ-Python | PIQE-MATLAB | CHILLUK-Python | FeatureNet-Python | Gabybaldeon-Python | RPIAXIS-Python | TeamEpoch-Python |
|---|---|---|---|---|---|---|---|---|---|
| OQ | 0.57 | 0.26 | 0.68 | 0.81 | 0.84 | 0.86 | 0.54 | 0.83 | 0.84 |
| ROI | 0.11 | 0.28 | 0.54 | 0.77 | 0.86 | | | 0.80 | 0.80 |
| NOISE | 0.90 | 0.57 | 0.62 | 0.74 | 0.37 | | | 0.82 | 0.89 |
| All Data with PW8-KV90 OIQ | 0.79 | 0.29 | 0.71 | 0.72 | 0.75 | 0.78 | 0.41 | 0.74 | 0.81 |
| All Data with PW8-KV90 ROI | 0.41 | 0.27 | 0.44 | 0.71 | 0.78 | | | 0.73 | 0.74 |
| All Data with PW8-KV90-NOISE | 0.81 | 0.46 | 0.57 | 0.65 | 0.25 | | | 0.75 | 0.82 |

**Table 6.** Spearman's rank correlation coefficient (SRCC) values, computed across the combined imaging datasets encompassing all phantoms and image acquisition settings, between each of the 27 calculated image quality assessment (IQA) measures and the expert ratings for overall image quality (OIQ), region-of-interest (ROI) image quality, and noise patch analysis. In addition, SRCC values are reported for OIQ, ROI and noise analyses performed separately for a subset with fixed kV and pulse width (PW8-KV90=pulse width 8 and kV 90).

## IQA measure–based rating

Since our experiments as outlined in previous sections showed that observers could not detect subtle image quality changes with varying kV and pulse width, also correlation between the annotators and IQA measures values were so low in those cases, we conducted an experiment to assess whether IQA measures could capture these subtle variations. For this purpose, we calculated IQA measure values on the OIQ and examined whether different kV and pulse width combinations were consistently ranked from highest to lowest quality among different IQA measures (Table 7). The results showed all IQA measures agreed and identified PW6.4-KV90 combination as the best quality, followed by PW5-KV90 and PW3.2-KV90, confirming their ability to capture subtle quality changes missed by manual ratings. This agreement suggests that image quality measures could quantify some changes in image quality which are not apparent to human eyes. Results on the ROI also confirmed this agreement across measures (Table 8). To overcome the limitation of human rating we therefore introduce these rankings as an IQA measure–based rating alongside manual observer ratings, as they may provide a valuable benchmark for developing new IQA measures and can help developer to ensure their new measures can capture such subtle image quality differences as established available measures.
We would like to mention that the IQA measure–based rating is exploratory. While the high level of agreement among IQA measures demonstrates their sensitivity to subtle image quality differences, it does not necessarily indicate a closer correspondence to diagnostic image quality than expert reader assessments. Rather, these results suggest that IQA measures capture image characteristics that are not readily perceived by human observers, enabling discrimination between images that received identical human quality ratings.

| | PW8-KV60 OIQ | PW8-KV125 OIQ | PW5-KV90 OIQ | PW6.4-KV90 OIQ | PW3.2-KV90 OIQ |
|---|---|---|---|---|---|
| Score | 1 | 1 | 3 | 4 | 2 |
| Full reference-IQA | | | | | |
| SSIM-matlab | 0,877 | 0,859 | 0,895 | 0,896 | 0,880 |
| MS-SSIM-matlab | 0,954 | 0,952 | 0,965 | 0,966 | 0,959 |
| IW-SSIM-matlab | 0,753 | 0,764 | 0,805 | 0,806 | 0,774 |
| CW-SSIM-matlab | 0,992 | 0,994 | 1,000 | 1,000 | 1,000 |
| PSNR-matlab | 32,212 | 31,728 | 37,163 | 37,304 | 36,248 |
| RMSE-matlab | 2,140 | 2,023 | 0,942 | 0,918 | 1,042 |
| DISTS-matlab | 0,056 | 0,048 | 0,031 | 0,030 | 0,034 |
| DISTS-python | 0,057 | 0,049 | 0,031 | 0,030 | 0,034 |
| **DSS-matlab** | 0,579 | 0,573 | 0,637 | 0,640 | 0,600 |
| **FSIM-matlab** | 0,960 | 0,960 | 0,972 | 0,973 | 0,967 |
| GMSD-matlab | 0,053 | 0,056 | 0,044 | 0,043 | 0,049 |
| **HaarPSI-matlab/python** | 0,810 | 0,809 | 0,854 | 0,858 | 0,834 |
| **HaarPSImed-python** | 0,676 | 0,668 | 0,721 | 0,726 | 0,692 |
| LPIPS-python | 0,040 | 0,034 | 0,022 | 0,021 | 0,024 |
| MDSI-matlab | 0,287 | 0,290 | 0,272 | 0,270 | 0,285 |
| **VIF-python** | 1,258 | 0,696 | 0,934 | 0,937 | 0,921 |
| **VSI-matlab** | 0,989 | 0,991 | 0,993 | 0,994 | 0,992 |
| AMBE-matlab | 636,805 | 1031,156 | 48,440 | 25,967 | 64,105 |

**Table 7.** Image quality assessment (IQA) measure values and overall image quality (OIQ) scores demonstrating strong agreement between the IQA measures in scoring different imaging settings, capturing subtle image quality differences that were not detected by the expert manual ratings.

| | **PW8-KV60 ROI** | **PW8-KV125 ROI** | **PW5-KV90 ROI** | **PW6.4-KV90 ROI** | **PW3.2-KV90 ROI** |
|---|---|---|---|---|---|
| Score | 1 | 1 | 3 | 4 | 2 |
| FR-IQA | | | | | |
| SSIM-matlab | 0,812 | 0,806 | 0,826 | 0,827 | 0,813 |
| MS-SSIM-matlab | 0,920 | 0,927 | 0,939 | 0,939 | 0,934 |
| IW-SSIM-matlab | | | | | |
| CW-SSIM-matlab | | | | | |
| PSNR-matlab | 29,058 | 30,110 | 34,198 | 34,268 | 33,626 |
| RMSE-matlab | 6,543 | 6,846 | 4,828 | 4,731 | 5,085 |
| DISTS-matlab | 0,107 | 0,084 | 0,066 | 0,066 | 0,068 |
| DISTS-python | 0,108 | 0,085 | 0,068 | 0,068 | 0,070 |
| **DSS-matlab** | 0,660 | 0,663 | 0,696 | 0,698 | 0,682 |
| **FSIM-matlab** | 0,922 | 0,934 | 0,946 | 0,947 | 0,942 |
| GMSD-matlab | 0,068 | 0,059 | 0,057 | 0,057 | 0,062 |
| **HaarPSI-matlab/python** | 0,756 | 0,787 | 0,808 | 0,809 | 0,792 |
| **HaarPSImed-python** | 0,615 | 0,636 | 0,659 | 0,661 | 0,640 |
| LPIPS-python | 0,054 | 0,039 | 0,025 | 0,025 | 0,027 |
| MDSI-matlab | 0,335 | 0,337 | 0,331 | 0,331 | 0,337 |
| **VIF-python** | 1,475 | 0,630 | 0,915 | 0,918 | 0,905 |
| **VSI-matlab** | 0,979 | 0,983 | 0,986 | 0,986 | 0,984 |
| AMBE-matlab | 833,829 | 1233,545 | 63,832 | 35,379 | 75,366 |

**Table 8.** Image quality assessment (IQA) measure values and Region of interest (ROI) scores demonstrating strong agreement between the IQA measures in scoring different imaging settings, capturing subtle image quality differences that were not detected by the expert manual ratings.

## Data Records

The dataset consists of 1764 slides related to CBCT images for different kV, pulse width and reconstruction types. The full data is available on Zenodo [49]. The python library nibabel was used to convert the DICOM files to NIfTI format to guarantee ease of use of our phantom scans with different software tools. This repository consists of the files named “degraded.zip” and “reference.zip” which contain the degraded and reference images saved in the NIfTI format. These files are named in the following format:

”./DATASETNAME/SLICE/KERNEL/KV/PULSEWIDTH/RECONSTRUCTIONMETHOD.nii.gz”.

For every degraded image in the dataset DATASETNAME and in slice SLICE, the corresponding reference image is stored under the path ”./DATASETNAME/SLICE/EE/090/8.0/NORMAL.nii.gz” in the “reference.zip” file. This means that for each slice, the image with kernel EE, kV 90, pulse width 8.0 and reconstruction type NORMAL is selected as reference image.

Furthermore, the file “annotations.csv” contains the annotations by three human experts for the tasks OIQ (ROI) in the columns OIQ-1, OIQ-2, OIQ-3 (ROI-1, ROI-2, ROI-3) and the mean of the zscored annotations in the column OIQ_MOS_ZSCORE (ROI_MOS_ZSCORE). Furthermore, the IQA measure-based rating for OIQ and ROI is stored in the columns “OIQ_IQA-rating“ and “ROI_IQA-rating“. The file „software_versions.txt“ contains the versions of the employed IQA measures.

## Technical Validation

The CBCT scanner within the research-dedicated Hybrid Theatre at the University of Sydney is not used clinically. However, it undergoes regular maintenance and calibration performed by the vendor in line with clinical sites. The CBCT acquisitions protocols were conducted by an experienced medical imaging researcher who routinely works with the scanner.

## Usage Notes

The imaging dataset will be publicly available through the Zenodo repository [40], together with detailed documentation describing the dataset structure, file organization, and recommended usage. All CBCT images are provided in NIfTI format, allowing straightforward visualization and analysis using widely available open-source medical imaging software, including 3D Slicer, Analyze 7.5, and ITK-SNAP. The expert annotations are provided in CSV format and can be readily accessed using diverse tools.

The dataset was specifically designed to support IQA research and benchmarking by providing systematically acquired CBCT images with controlled variations in imaging parameters, including kV, PW and RT. For every image, corresponding expert image quality ratings are available for OIQ, ROI image quality.

To facilitate immediate use and ensure reproducible benchmarking, we additionally provide the computed values of 26 established IQA measures, including both full-reference and no-reference (NR) methods, together with references to the original implementations used for their computation. Furthermore, we introduce an IQA measure–based consensus rating that summarizes the relative image quality across acquisition settings based on the most reliable objective measures identified in this study. These resources enable researchers to benchmark newly developed IQA methods against established measures, evaluate the robustness and consistency of IQA algorithms under varying acquisition conditions, investigate agreement between objective measures and expert visual assessment, and compare the performance of different IQA approaches using a common reference dataset.

Potential applications of the dataset include the development and validation of novel IQA algorithms, benchmarking of full-reference and no-reference IQA methods, optimization of CBCT acquisition protocols, objective evaluation of reconstruction algorithms, AI-based image quality prediction, quality-aware image enhancement, automated quality control systems for CBCT imaging, and studies investigating the relationship between objective image quality measures and human perception.

**Data Availability**

The data are published on Zenodo [49].

**Code Availability**

The software versions of the used IQA measure implementations are listed on Zenodo [49] in the file “software_versions.txt”.

**Acknowledgment**

This study was supported by NÖ FTI Grundlagenforschung project (Project number: GLF21-1-001). In addition, this study was funded by ACMIT COMET Module FFG project (FFG number: 879733, application number: 39955962). We also acknowledge funding from the Austrian Science Fund (FWF) through project T1307. Furthermore, this study was also supported by a University of Sydney Robinson Fellowship and an NHMRC Investigator Grant (APP ID: 1041194).

**Author Contribution**

SH, ABr, CK, ABi, TR conceptualized the study, the same authors and PM, AAA, PA, WB, AP, GK, CBS have been involved in methodological development and evaluation, TR performed all the measurements, SH, ABr, CK did the analysis of the whole data, CK conducted the experimental computation, SH, ABr supervised the study design, SH provided the first paper draft and all authors contribute in writing and revising the paper.

**Competing Interests**

The authors declare no competing interests.